# Development of Ni doped SnO$_2$ Dilute magnetic oxides for electronics and spintronics applications

Y.S. Worku [a], V. V. Srinivasu[a], Dipti R. Sahu[b*]

[a] Department of Physics, University of South Africa, Florida science campus, Johannesburg, South Africa
[b] Department of Applied Sciences, Namibia University of Science and Technology, Windhoek, Namibia

**Abstract**

We used a solid-state reaction method to prepare $Sn_{1-x}Ni_xO_2$ with x = 0, 0.05, 0.1, 0.15 polycrystalline compounds. A rutile phase with tetragonal crystal structure was confirmed by X-ray diffraction. At room temperature, the magnetisation study shows that the saturation magnetization increases with Ni doping content whereas the coercive field decreases after x= 0.1. The spin number increases as the Ni doping concentration increases, indicating that the incorporation of Ni into the Sn sites increases the number of spins interacting to improve the ferromagnetic phase, which is like saturation magnetization and coercive field.

## 1. Introduction

In recent years, dilute magnetic semiconductors have gained the interest of the scientific community due to their use in spintronics applications [1, 2] . According to recent experimental findings the existence of ferromagnetism at room temperature in Ni doped SnO$_2$ has made it promising material for spintronics applications [3]. However, the cause of ferromagnetism properties in these materials is not clearly understood and whether the ferromagnetism is originates from intrinsic property or extrinsic factors is not certain. Few studies have demonstrated that doping SnO$_2$ with Mn does not show ferromagnetism [2,4]. On the other side Cr:SnO$_2$, Al: SnO$_2$, Fe:SnO$_2$, Co:SnO$_2$ and V:SnO$_2$ exhibit ferromagnetism at room temperature [2,5,6,7]**.** There are reports showing oxygen vacancies and structural defects might be the possible cause for the establishment of ferromagnetism [2-7]. In some situations the addition of more dopants Ni ions can bring the magnetic moment of each Ni ion to decrease, which is attributed to antiferromagnetic super-exchange interactions between Ni ions in the system [3,7]. In this study we study the magnetic behaviour in Ni doped SnO$_2$ materials with different Ni content for possible spintronics application.

## 2. Experimental Method

The polycrystalline powder samples of $Sn_{1-x}Ni_xO_2$ with x = 0, 0.05, 0.1, 0.15 were prepared by the slow state solid state reaction using SnO$_2$ and NiO. The calcination and presintering of the samples were carried out at 200 °C and 400 °C for 8 h at each temperature. The final sintering of the samples was performed at 800 °C for 16 h in pellet form. X-ray diffraction (XRD) was carried out for structural analysis using Philips diffractometer (Model 1715). Magnetic measurements were carried out using superconducting quantum interference device (SQUID) magnetometer. The spin system of $Sn_{1-x}Ni_xO_2$ is examined using electron spin resonance (ESR) spectroscopy.

---

*Corresponding author
Email: dsahu@nust.na (D. R. Sahu)

### 3. Results and discussion

The X-ray diffraction patterns indicates tetragonal rutile-type crystal structure [JCPDS card no. 14- 4145] as shown in Fig. 1.

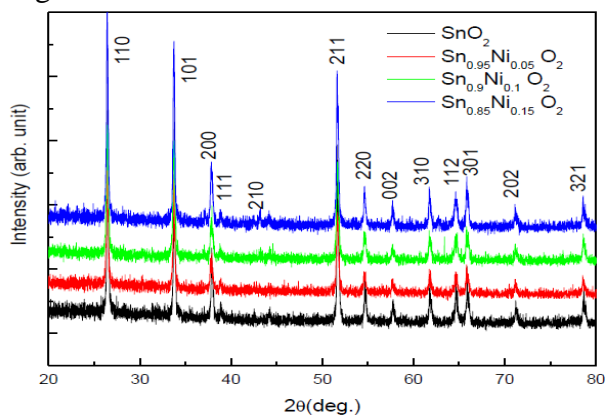

Figure 1: X-ray diffraction patterns of $Sn_{1-x}Ni_xO_2$ (x = 0, 0.05,0.1 0.15) compound.

The average crystal size for $Sn_{1-x}Ni_xO_2$ (x = 0, 0.05, 0.1, 0.15) particles can be obtained utilizing Debye Scherrer's formula (Dv) and from the Debye formula we found that increaing Ni doping content in the $SnO_2$ lattice causes the crystallite size to reduce. This indicates $Ni^{2+}$ ion replacement disrupts the $SnO_2$ lattice, resulting in lattice deformation and a reduction in crystallite size [6]. The TEM images for $SnO_2$ and $Sn_{0.9}Ni_{0.1}O_2$ particles are shown in Fig. 2. The images shows that grain becomes more spherical and small. This happens as Ni radius is smaller than Sn atom resulting in a decrease of particle size inducing vlume reduction due to the restricted solubility and surface segregation of the solid solution [7]. The magnetization study indicates that saturation magnetization increases with Ni doping content [8], whereas coercivity field ($H_C$) rise first up to x = 0.1 of Ni doping and there after drop as Ni doping increases. A deviation from a linear relation between magnetization and Ni concentration is indicated by the presence of antiferromagnetic interactions between Ni ions through O atoms resulting in the observed decline in coercivity value [9].

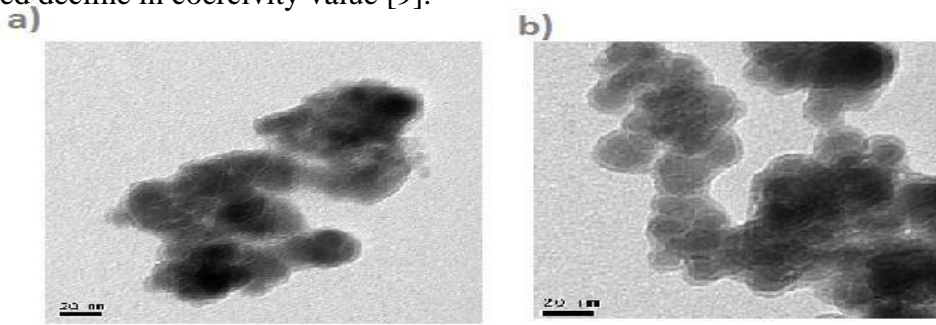

Figure 2: TEM micrographs of (a) pure $SnO_2$ and (b) $Sn_{1-x}Ni_xO_2$ for x=0.

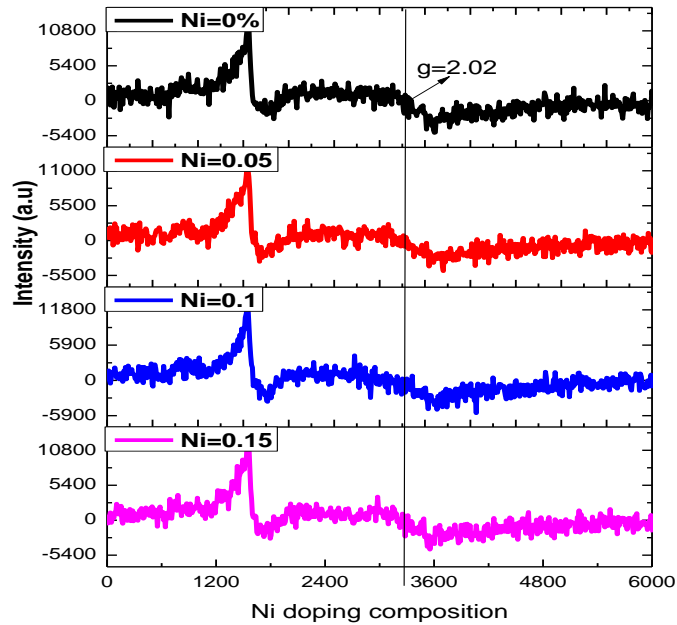

Figure 4: Room temperature ESR spectra for $Sn_{1-x}Ni_xO_2$ (x = 0, 0.05, 0.1 0.15).

The spin system of $Sn_{1-x}Ni_xO_2$ nano-crystalline is examined using electron spin resonance (ESR) spectroscopy. As shown in Fig. 4 the value of g factor equals to g =2.02 which is interpret as an oxygen vacancy trapped in $Ni^{2+}$ producing $Ni^{3+}$ that $Ni^{+3}$ ions interact via oxygen vacancy to create ferromagnetim. The total number of spins (N) involved in ESR spectra was calculated using the equation N=0.285( $I_{P-P}$ ) $(\Delta H)^2$ where Ip-p is the peak-to-peak height of the ESR signal measured from the center of the origin, ΔH is the line width in Gauss units [10]. The spin number increases with increasing of Ni doping content between Ni content ($0 \leq x \leq 0.1$) this is due to the incorporation of Ni into the Sn sites increases the number of spins interacting to enhance ferromagnetic phase. Further Ni doping at x=0.15 decreases spin number which might be due to antiferromagnetic interaction indicates that both extrinsic and intrinsic factors are ultimately responsible for the magnetic behavior of this Ni doped $SnO_2$ system.

## 4. Conclusion

Ni doped $SnO_2$ synthesized using slow step solid state sintering technique shows rutile-type tetragonal structure. Magnetization study indicates coercive field declines after 0.1 of Ni doping content. ESR spectra results corroborates that the number of spins interacting for the formation of ferromagnetic is similar to the result of coercivity versus Ni doping content. This result implies that good dilute magnetic semiconductors with increased magnetic behavior may be developed by optimizing the Ni percent doping ratio for spintronics applications.

### Acknowledgment

The authors thank University of South Africa and Namibia University of Science and Technology for funding this research.